\documentclass[usenatbib]{mn2e}
\usepackage{graphicx}
\usepackage{astrobib_mnras2e}

\begin{document}
\topmargin-1cm

\def\bi#1{\hbox{\boldmath{$#1$}}}

\newcommand{\be}{\begin{equation}}
\newcommand{\ee}{\end{equation}}
\newcommand{\bea}{\begin{eqnarray}}
\newcommand{\eea}{\end{eqnarray}}

\newcommand{\lexp}{\mathop{\langle}}
\newcommand{\rexp}{\mathop{\rangle}}
\newcommand{\rexpc}{\mathop{\rangle_c}}

\def\bi#1{\hbox{\boldmath{$#1$}}}

\title{Stellar and Dynamical Masses of Ellipticals in the Sloan Digital Sky Survey}

\author[Padmanabhan et al]
{Nikhil Padmanabhan$^{1}$\thanks{npadmana@princeton.edu}, Uro\v s Seljak$^{1}$, Michael A. Strauss$^{2}$,
Michael R. Blanton$^{3}$, \and
Guinevere Kauffmann$^{4}$, David J. Schlegel$^{2}$,
Christy Tremonti$^{5}$, 
Neta A. Bahcall$^{2}$, \and
Mariangela Bernardi$^{6}$, J. Brinkmann$^{7}$, 
Masataka Fukugita$^{8}$, \v{Z}eljko Ivezi\'{c}$^{2}$ \\
$^{1}$Joseph Henry Laboratories, Jadwin Hall, Princeton University, Princeton, NJ 08544 \\
$^{2}$Dept. of Astrophysical Sciences, Peyton Hall, Princeton University, Princeton, NJ 08544 \\ 
$^{3}$Dept. of Physics, NYU, 4 Washington Place, New York, NY 10003 \\
$^{4}$Max Planck Institut fur Astrophysik, Karl-Schwarzschild-Str. 1, D-85741 Garching, Germany \\
$^{5}$Steward Observatory, 933 N Cherry Ave., Tucson AZ 85721-0065 \\
$^{6}$Department of Physics, Carnegie Mellon University, 5000 Forbes Avenue, Pittsburgh, PA 15213 \\
$^{7}$Apache Point Observatory, P.O.Box 59, Sunspot, NM 88349-0059 \\
$^{8}$Institute for Cosmic-Ray Research, University of Tokyo, Kashiwa 277-8582, Japan 
}

\maketitle

\begin{abstract}
We study the variation of the dark matter mass fraction of elliptical galaxies as a 
function of their luminosity, stellar mass, and size using a sample of 
29,469 elliptical galaxies culled from the Sloan Digital Sky Survey. We 
model ellipticals as a stellar Hernquist profile embedded in an adiabatically
compressed dark matter halo. This model
allows us to estimate a dynamical mass ($M_{dynm}$) at the half-light radius
from the 
velocity dispersion of the spectra, and to compare these to the stellar
mass estimates ($M_{*}$) from \cite{2003MNRAS.341...33K}. 
We find that $M_{*}/L$ is independent of luminosity, while 
$M_{dynm}/L$ increases with luminosity, implying that
the dark matter fraction increases with luminosity.
We also 
observe that at a fixed luminosity or stellar mass, 
the dark matter fraction increases with increasing galaxy size 
or, equivalently, increases with decreasing surface brightness:
high surface brightness galaxies show almost no evidence for
dark matter, while in low surface brightness galaxies, the  
dark matter exceeds the stellar mass at the half light radius.
We relate this to the fundamental plane 
of elliptical galaxies, suggesting that the tilt of this plane from 
simple virial predictions is due to the dark matter in galaxies. 
We find that a simple model where galaxies are embedded in 
dark matter halos and have a star formation efficiency
independent of their surface brightness explains these trends.
We estimate the virial mass
of ellipticals as being approximately 7-30 times their stellar mass, 
with the lower limit suggesting almost all of the gas within the 
virial radius is converted into stars. 
\end{abstract}



\section{Introduction}

The zoo of galaxies is an eclectic one, with large variations in morphologies,
colours and spectra. This is perhaps not very surprising, given the varied 
environments in which galaxies reside and the different processes underlying
their formation. However, 
a number of regular trends are also known; for instance, 
the rotation velocities of spiral galaxies and the central velocity
dispersions of elliptical galaxies correlate strongly with luminosity -- the
\cite{1977A&A....54..661T} and \cite{1976ApJ...204..668F} relations.
In addition, elliptical galaxies appear to reside in a thin plane in the 
space of their luminosity ($L$), radius ($R$), and central velocity
dispersion ($\sigma$), the ``fundamental plane'' 
\citep{1987ApJ...313...59D,1987ApJ...313...42D}, of which 
the Faber-Jackson relation is a projection.
Understanding the source of these regularities 
is still an open problem; they provide a strong constraint and challenge to
theories of galaxy formation.

Elliptical galaxies are probably the simplest galaxies to
model. These are pressure supported systems, and are remarkably uniform in
their properties. Their intensity profiles are well described by a  
\cite{1948AnAp...11..247D} profile, $I \propto \exp(-(\theta/\theta_{s})^{1/4})$, they have very 
simple spectral energy distributions and they have very uniform photometric properties.
However, very little has been known about the matter content of elliptical
galaxies until recently, principally due to the lack of dynamical tracers like
the $H I$ rotation curves for spiral galaxies.
More recently however, there have been a number of surveys that have measured 
the dynamical structure of samples of elliptical galaxies using 
slit spectroscopy \citep{2000A&AS..144...53K},
and with the advent of integral field spectroscopy \citep{2002MNRAS.329..513D}, the
number of well studied elliptical galaxies will only increase. In addition,
strong lensing measurements \citep[eg.][]{2003ApJ...583..606K} 
have added independent measures of the masses of
the halos of ellipticals, especially at radii outside the 
realm of spectroscopic techniques.

The Sloan Digital Sky Survey \cite[SDSS,][]{2000AJ....120.1579Y}
presents a different model for studying
elliptical galaxies. Unlike the surveys above that collect very detailed 
dynamical information for a small number of galaxies, the SDSS, with its $\pi$ steradians
of deep multicolour imaging and $10^{6}$ spectra, will measure global parameters
for a very large number of galaxies, allowing one to 
statistically approach questions of galaxy structure. Indeed, the properties of elliptical
galaxies in the SDSS have already been studied in great detail 
(Bernardi et al \citeyear{2003AJ....125.1817B}, \citeyear{2003AJ....125.1849B}, 
\citeyear{2003AJ....125.1866B}, \citeyear{2003AJ....125.1882B}).
This work constructed a sample of 9000 galaxies from early SDSS data, and 
exhaustively measured the various correlations between different observables. Principal 
results include a Faber-Jackson relation between luminosity $L$ and 
velocity dispersion $\sigma$, $\sigma \propto L^{0.25 \pm 0.012}$,
and a fundamental plane relation between the effective radius $R$, 
the half-light surface 
brightness $I$, and $\sigma$, $R \propto
\sigma^{1.49\pm0.05} I^{-0.75\pm0.01}$.
This statistical approach has been further extended using the 
weak lensing information from galaxy-galaxy lensing at large radii
\citep{2001astro.ph..8013M} and models satisfying the dynamical 
constraints from velocity dispersions and weak lensing in an average sense
have been constructed \citep{2002MNRAS.334..797S}.

This paper explores
the dark matter content of elliptical galaxies in the SDSS in more 
detail.
Is there evidence for dark matter
in ellipticals? How does it correlate with the luminosity and size
of the galaxy? How does the dynamical mass compare to
the stellar mass of elliptical galaxies? The simplest virial prediction combined
with a constant mass to light ratio implies that $L \propto \sigma^{2} R$; however,
the observed fundamental plane (FP) of ellipticals shows the scaling
$L \propto \sigma^{1.98} R^{0.66}$ 
\citep{2003AJ....125.1866B}. There have been a number of proposals in the literature
to explain this deviation from the virial prediction, or ``tilt''. Implicit in the 
virial prediction is the assumption that elliptical galaxies form a homologous sequence, i.e.
the dynamical structure of elliptical galaxies are self similar and related by simple scaling relations.
Deviations from such a sequence would naturally manifest themselves as a tilt in the FP.
However, based on detailed dynamical measurements of nearby ellipticals, 
\cite{2001AJ....121.1936G} argue that there is little evidence for any deviations
from homology. A second, and preferred, proposal is to assume that $M/L$ is not constant, 
but varies with luminosity. Such a variation could be caused either due to a metallicity (or dust)-
luminosity correlation, or due to an increase in the dark matter fraction with luminosity. Is it
possible to distinguish between these scenarios?
These are the questions that 
this paper will attempt to answer.

We start in Section 2 with the criteria for selecting our sample of SDSS elliptical galaxies.
In Section 3, we propose a mass model for these galaxies and use it to
estimate dynamical masses based on measured scale sizes and velocity dispersions. 
We then compare stellar and dynamical mass estimates 
as a function of galaxy properties and discuss their implications (Sections 4 and 5). The appendix
summarizes the relevant properties of the Sersic profiles we use. Wherever needed, this paper uses 
($\Omega_{m},\Omega_{\Lambda}$) = (0.3,0.7) and $H_{0} = 70$ km/s/Mpc.

\section{The Sample}

The SDSS is imaging 10$^{4}$ deg$^{2}$ of the Northern Galactic Cap in 5 bandpasses 
\citep[$u,g,r,i,z$]{1996AJ....111.1748F} using a drift scanning, mosaic CCD camera
\citep{1998AJ....116.3040G} under photometric conditions \citep{2001AJ....122.2129H},
and is targeting 10$^{6}$ objects for spectroscopy \citep{2003AJ....125.2276B}, 
most of which are galaxies with $r$ band
apparent magnitude $m_{r} <$ 
17.77 \citep{2002AJ....124.1810S}. 
The data used in this paper cover an area of $\sim$ 2000 deg$^{2}$ and include $\sim$
160,000 galaxies with spectra, and are denoted
{\bf sample10} within the SDSS collaboration.
All of these data have been reduced by highly automated photometric and
spectroscopic reduction pipelines 
\cite[see][for details]{2002AJ....123..485S}. The astrometric calibration 
is automatically performed by a pipeline that obtains absolute positions to better
than 0.1 arcsec \citep{2003AJ....125.1559P}, and magnitudes are calibrated to a standard star network
approximately in the AB system \citep{2002AJ....123.2121S}.

The SDSS pipelines return a wealth of information for all detected objects; in addition,
a number of auxiliary parameters have been measured by 
various members of the SDSS collaboration for these objects. The principal parameters
relevant to this work are mentioned below.
\begin{itemize}
\item Redshifts($z$) and Velocity dispersions($\sigma$): Each of the SDSS galaxy spectra is fit to a 
linear combination of galaxy templates at varying redshifts, broadened by a 
Gaussian kernel (Schlegel et al. \citeyear{schlegel}). Minimizing $\chi^{2}$ over this 
suite of models leads to an estimate of both the redshift and the stellar velocity 
dispersion of the galaxy.
\item Petrosian Fluxes : The primary measure of galaxy flux in the SDSS is the
Petrosian magnitude, a modification of the quantity defined by \cite{1976ApJ...209L...1P}; see 
\cite{2002AJ....124.1810S} for details. 
Note that in the absence of seeing, the Petrosian flux is about 
81.5\% of the total flux of a deVaucouleurs profile.
The photometric pipeline also returns the radii that enclose
50\% and 90\% of the Petrosian flux, $R_{p,50}$ and $R_{p,90}$ respectively, as well
as the ratio of the minor to major axes, $a/b$, of the galaxy from model fits to
a deVaucouleurs profile.
\item K-corrections : All luminosities and colours used in this paper are k-corrected
to the median redshift of the survey, $z_{med} = 0.1$, using the $kcorrect$ package 
of \cite{2003AJ....125.2348B}.
\item Sersic Profiles : The radial intensity profiles of
all of the galaxies in our sample have been fit to Sersic 
profiles \citep{1968AGalA.B...0000S,2002astro.ph..9479B},
\be
I(R) = A \exp \left[ - (\theta/\theta_{s})^{1/n} \right] \,\,,
\ee
where $R_{s}$ is a scale radius, and $n$ measures the concentration of the 
intensity profile \footnote{We summarize the relevant properties of the 
Sersic profile in the Appendix.}. Unlike the Petrosian magnitudes, the Sersic 
fits are convolved with the PSF. Therefore, the effective or half light 
radius of the Sersic profile, $R_{50}$, is a robust estimator of the 
galaxy size and is what we adopt throughout this paper.
\item $D_{4000}$: The continuum break at 4000 \AA, $D_{4000}$ \citep{1999ApJ...527...54B}
is one of the most prominent
features in galaxy spectra, and is caused by the accumulation of a large
number of metal lines in a narrow region of the spectrum. Hot stars, indicative of a young stellar
population, show a weak 4000 \AA\, break as the principal metals causing the 
absorption are multiply ionized, making this feature a powerful age estimator.
The strength
of the break is computed following \cite{tremonti} \cite[see also][]{2003MNRAS.341...33K}.
\item Stellar masses ($M_{*}$): \cite{2003MNRAS.341...33K} compute stellar mass to light ratios for all
spectroscopically observed galaxies by fitting stellar population synthesis models to 
the measured $D_{4000}$ strength and the $H\delta_{A}$ absorption line, assuming the \cite{2001MNRAS.322..231K}
initial mass function (IMF). Given this $M/L$ ratio, the stellar mass is estimated by
multiplying by the Petrosian luminosity in the $z$ band, correcting for extinction due to dust by matching 
the predicted galaxy colours to the observed colours. 
\end{itemize} 

\begin{figure}
\begin{center}
\leavevmode
\includegraphics[width=3.0in]{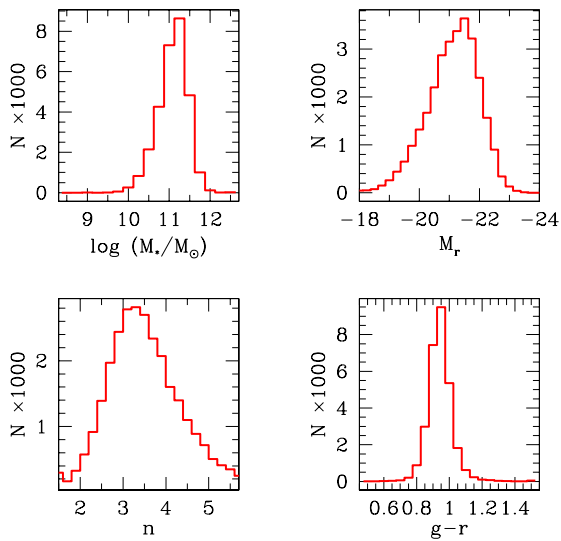}
\end{center}
\caption{Properties of the sample of elliptical galaxies. The upper left panel 
shows the distribution of the logarithm of stellar masses, while the upper
right panel is the distribution of the $r$-band absolute magnitudes. The lower left panel shows 
the Sersic indices for our sample, where n = 4 is the de Vaucouleurs profile, and the
lower right panel is the distribution of the g-r k-corrected colour. 
}
\label{fig:sample}
\end{figure}

\subsection{Selecting Ellipticals}

In order to translate the velocity dispersions into dynamical mass estimates, it is 
important that we select galaxies that are
pressure supported, and not rotationally supported. In 
practice, this closely corresponds to selecting a sample of elliptical galaxies.
We  note that our selection criteria will also allow S0's; these galaxies, although
not strictly elliptical galaxies, satisfy our criterion that they have little or 
no rotational support. However, for simplicity, we refer to all these galaxies as
elliptical galaxies.
Since we desire minimal contamination from spirals, we use a combination of 
spectroscopic and photometric criteria,
similar in spirit to those in \cite{2003AJ....125.1817B}, although differing in detail. 
Note that unless specified, all the photometric quantities are measured in the $r$ band.
The cuts made were 
\begin{itemize}
\item $D_{4000} > 1.6$ : As mentioned above, a large $D_{4000}$ corresponds to an 
older stellar population, normally associated with ellipticals. This particular
cut is based on \cite{2003MNRAS.341...33K}.
\item $R_{p,90}/R_{p,50} > 2.6$ \citep{2001AJ....122.1238S}: 
Elliptical galaxies have intensity profiles that are 
more concentrated than spirals, allowing one to separate them on the basis of a 
concentration parameter, $R_{p,90}/R_{p,50}$.
\item No emission lines : The stellar populations of ellipticals are generally old, with
little or no star formation. Since emission lines are associated with star forming regions,
we reject galaxies with emission lines ($H\alpha, NII$) detected with S/N $>$ 3.
\item $\sigma > 70$ km/s : This cut is based on the resolution of the SDSS spectra; velocity
dispersions lower than 70 km/s cannot be reliably determined.
\item Minor/Major axis ratio ($a/b$) $>$ 0.7 : This was chosen to eliminate large
edge-on spiral galaxies that were not eliminated by the
previous cuts. Also, since a large rotation velocity would tend to
flatten elliptical galaxies, this eliminates galaxies with large rotation velocities. 

\end{itemize}

Making these cuts reduces our sample from 165,812 to 29,469 galaxies. Fig. \ref{fig:sample} 
shows some of the properties of our sample.
Most of our galaxies have Sersic index $n$ between 3 and 4 and have 
their Petrosian $g-r$ colour narrowly distributed around $\sim$ 1.0.
We did not cut on either of these quantities, yet they lie in the range expected for elliptical
galaxies.
In addition, we have visually
inspected images of a random subsample of these galaxies and have found the contamination
from spirals to 
be less than 5 $\%$. 

\section{A Dynamical Model}
\label{sec:model}

\begin{figure}
\begin{center}
\leavevmode
\includegraphics[width=3.0in]{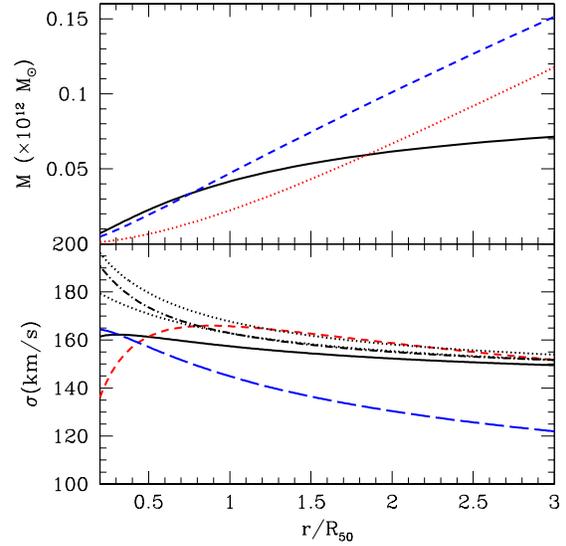}
\end{center}
\caption{ The upper panel shows the integrated mass of the original
matter (dotted/red) profile, the stellar (solid/black) and final 
dark matter (dashed/blue) distributions. The lower panel is the 
luminosity weighted velocity dispersion profile for an
isotropic $\beta=0$ (solid), $\beta = 0.3$ (lower dotted) and 
$\beta = 0.5$ (upper dotted)
velocity distribution. We also show a variable anisotropy profile (dot-dashed),
$\beta = 4\beta_{0}x/(1+x)^{2}$ where $x = r/r_{\beta}$, and $\beta_{0}=0.5$
and $r_{\beta}=$ 1.38 kpc. The short dashed
(red) line is the circular velocity profile reduced by 1.65, while the 
long dashed (blue) line is the profile that would be 
observed by a slit spectrograph (using $\beta=0$). Both panels 
assumed a galaxy halo of 10$^{12} M_{\odot}$ with a concentration $(r_{virial}/r_{s})$
of 10, and a stellar component of 10$^{11} M_{\odot}$ with an 
effective radius of 5 kpc.
}
\label{fig:model}
\end{figure}

In order to relate the velocity dispersions to dynamical masses, we 
need to develop a model of the mass distribution of the galaxy. We model
each galaxy (Fig.\ref{fig:model}, top panel)
as having two spherically symmetric
components, a stellar component described by a \cite{1990ApJ...356..359H}
profile, 
\be
\nu \propto r^{-1} (r+a)^{-3} \,\,\, , 
\ee
and a dark matter halo. We adopt the
Hernquist profile as it provides a convenient analytical approximation to the 
deprojected deVaucouleurs profile \footnote{ See the Appendix for an
explanation of why this is 
still justified for the Sersic profile fits that we use.}.
Modelling the dark matter is more difficult since it
involves understanding its response to the baryons. We start with an NFW profile \citep{1997ApJ...490..493N} ,
\be
\rho \propto r^{-1} (r + r_{s})^{-2}\,\,\, ,
\label{eq:NFW}
\ee
for the initial matter 
distribution, assume that a fraction $F$ (the stellar mass fraction) condenses into
stars, and that the remaining matter (the dark matter mass fraction)
is adiabatically compressed by the 
stellar matter following \cite{1986ApJ...301...27B}. Determining the final
dark matter profile involves solving the equations
\bea
r_{i} M_{NFW}(r_{i}) = r_{f} (M_{*} (r_{f}) + M_{DM}(r_{f})) \nonumber \\
(1-F) M_{NFW}(r_{i}) = M_{DM}(r_{f}) \,\,\, ,
\eea
where $M_{NFW}(r_{i})$ is the initial mass profile, $M_{*}(r_{f})$ and $M_{DM}(r_{f})$ 
are the final
stellar and dark matter profiles, and $F$ is the stellar mass fraction. The first equation
represents angular momentum conservation, while the second enforces the fact that orbits 
of dark matter particles do not cross. 
We now use 
the Jeans equation to find the 3D velocity dispersion profile, $\overline{v^{2}_{r}}$ 
\citep{1987gady.book.....B},  
\be
v_{c}^{2}(r) \equiv \frac{G M(r)}{r} = -\overline{v^{2}_{r}} 
\left( \frac{d \ln \nu}{d \ln r} + \frac{d \ln \overline{v^{2}_{r}}}{d \ln r} + 2 \beta \right)\,\,\, ,
\ee
where $M(r)$ is the total (stellar and dark) 
mass within $r$, $v_{c}$ (the circular velocity) is the 
velocity a particle on a circular orbit would have, and 
$\beta$ measures the anisotropy of the velocity distribution,
\be
\beta = 1 - \frac{\overline{v_{\theta}^{2}}}{\overline{v_{r}^{2}}} \,\,\, .
\ee

While the applicability of this 
model to elliptical galaxies remains unclear, numerical 
simulations 
suggest it gives a good qualitative description in the region 
of interest $(\ga 0.5 R_{50})$
(M. Steinmetz, private communication). Fig. \ref{fig:model} (top panel) 
shows the initial and final integrated dark matter mass as a function of radius, as well as the stellar mass. 
The final circular velocity profile  shown in Fig. \ref{fig:model} (solid
line, bottom panel, the velocity was reduced by 1.65 as discussed below)
is nearly flat over the range of interest, in agreement with observations
of normal ellipticals \citep{2001AJ....121.1936G}. 
The reader is referred to \cite{2002MNRAS.334..797S} for more 
examples and the circular velocity profiles of individual components.

We then can project this distribution to two dimensions,
\be
I(R)\hat{\sigma^{2}}(R)  = 2 \int^{\infty}_{R} 
\left(1-\beta \frac{R^{2}}{r^{2}} \right)
\frac{\nu \overline{v^{2}_{r}}\, dr}{\sqrt{r^{2} - R^{2}}} \,\,\, ,
\ee
where $I(R)$ is the 2D intensity distribution. The SDSS spectrograph measures the 
luminosity weighted average (within the fiber aperture) of the above quantity,
\be
\sigma^{2}(R) = \frac{\int^{R}_{0} \,R'\, dR' \, I(R')\hat{\sigma^{2}}(R')}
{\int^{R}_{0} \,R'\, dR' \,I(R')}.
\ee
Examples of velocity dispersion profiles for
different values of $\beta$ are shown in the lower panel of Fig. \ref{fig:model}.

To compute the dynamical mass, we must choose a characteristic radius and relate
$\sigma$ at that radius to the circular velocity, $v_{c}$. The first complication is that 
the anisotropy $\beta$ is unknown; however, as Fig. \ref{fig:model} suggests, the effects of $\beta$
are more pronounced in the inner regions suggesting that we 
measure $\sigma$ at large radii. This is a manifestation of the 
virial theorem; for an isothermal profile, the luminosity
weighted velocity dispersion is simply related to the circular velocity via $v_c^2=3\sigma^2$.
From Fig. \ref{fig:model}, we see that $v_{c}(R_{50}) \approx 1.65 \sigma(R_{50})$ with a scatter 
of $\sim$ 10\%, depending of the anisotropy profile we use. This gives us a dynamical 
mass estimate at $R_{50}$,
\be
M_{dynm} = {(1.65 \sigma)^{2} R_{50} \over G}.  
\ee
This dynamical mass estimate is a 3D mass, while the stellar masses are projected masses.
We estimate the 3D stellar mass by considering the mass contained within the 
Hernquist profile at $R_{50}$, the projected half-light radius, which is
approximately $\sim 42\%$ of the total stellar mass.
Furthermore, the stellar masses are computed using the
Petrosian flux, which is approximately $80\%$ of the flux of an early type galaxy;
these two factors together imply that the 3D stellar mass at $R_{50}$ is $\sim 50\%$ of the
total estimated stellar mass.

\begin{figure}
\begin{center}
\leavevmode
\includegraphics[width=3.0in]{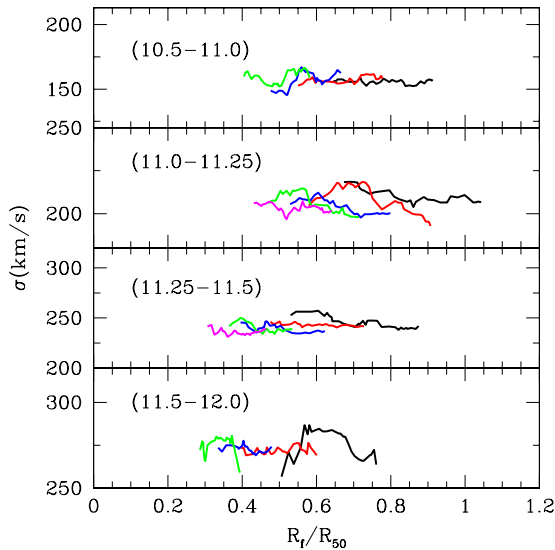}
\end{center}
\caption{ Estimated velocity profiles for our sample of ellipticals in SDSS as a
function of $R_{f}/R_{50}$ where $R_{f}$ is the fiber radius of 1.5 arcseconds. The
different panels are different stellar mass volume limited 
subsamples, where the panels are labelled by the logarithm of the stellar mass range.
The different 
line segments in each panel show the median velocity dispersion of subsamples in 
the physical size of the galaxy. Note that the size of the galaxy increases from
right to left. The various size ranges considered
are  (from top to bottom) 2.5-6.0, 3.0-6.0, 4.0-10.0
and 5.0-13.0 kpc.
The volume limited subsamples ensure a uniform sample over
the redshift range of interest.
The velocity profiles derived here are consistent with being flat over the
region of interest.
}
\label{fig:velprof_plot}
\end{figure}

Unfortunately, the velocity dispersion estimated by the SDSS spectra is not
at $R_{50}$, but at the fiber diameter of 3 arcseconds. One solution is to follow
\cite{2003AJ....125.1817B} and apply an empirical correction to the measured 
dispersion. Another approach, also suggested there, is to stack galaxies of similar 
masses and physical sizes and construct a velocity dispersion profile. These galaxies
are at a range of redshifts, so the 3 arcsecond fibers probe the velocity dispersion
at different physical radii, allowing us to construct a composite velocity profile.
We can then use this profile to correct the measured velocity dispersions 
out to $R_{50}$.

However, naively making subsamples is dangerous because the SDSS spectroscopic sample
is magnitude limited, i.e. to a good approximation, it included all galaxies with
$m_{r} < 17.77$. This implies that galaxies at higher redshifts are more luminous 
on average, which could introduce spurious correlations between observed quantities.
Therefore, to ensure uniformity,
all the subsamples we construct
in this paper are volume limited, including only those galaxies which would
remain in the sample at the redshift limits.

Fig. \ref{fig:velprof_plot} shows the velocity profiles estimated by stacking
galaxies. The different panels are different volume limited mass samples that
uniformly sample the redshift range, while the different line segments are the median
velocity dispersions of subsamples
of similar physical sizes. 
The profiles are consistent with being flat. This would appear inconsistent with the 
velocity dispersion profiles measured from individual ellipticals which, in general, 
decline with radius \citep[e.g.][]{2000A&AS..144...53K}. 
However, the profiles
are significantly flatter when luminosity weighted within the 
fiber, shown 
in Fig.\ref{fig:model} comparing the profile that would 
be seen through a slit with that through a fiber. We therefore use the 
velocity dispersion measured through the 3 arcsecond aperture as the 
dispersion at $R_{50}$. While the statistical velocity dispersion 
profiles are in a good agreement with the isotropic model predictions, 
the narrow dynamic range of volume limited subsamples prevents us
from making any strong conclusions on the anisotropy parameter $\beta$.

\section{Results}

\begin{figure}
\begin{center}
\leavevmode
\includegraphics[width=3.0in]{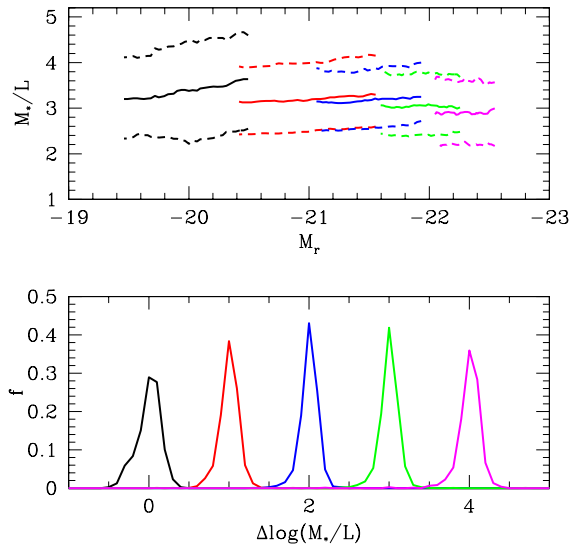}
\end{center}
\caption{The upper panel is the stellar mass to light ratio ($M_{*}/L$) as a function
of $r$ band luminosity. The different segments/colours correspond 
(from left to right) to volume limited subsamples
within redshift intervals 0.04-0.08 (black), 0.08-0.12 (red), 0.12-0.16 (blue), 
0.16-0.20 (green), 0.20-0.24 (magenta). The solid line is the median of the 
distribution, while the lower and upper dashed lines are the 16\% and 84\% intervals.
The lower panel shows the distribution of $M_{*}/L$ at the median luminosity of each
redshift subsample, staggered by $\Delta \log (M_{*}/L)=1$ for clarity.
}
\label{fig:logml_l}
\end{figure}

\begin{figure}
\begin{center}
\leavevmode
\includegraphics[width=3.0in]{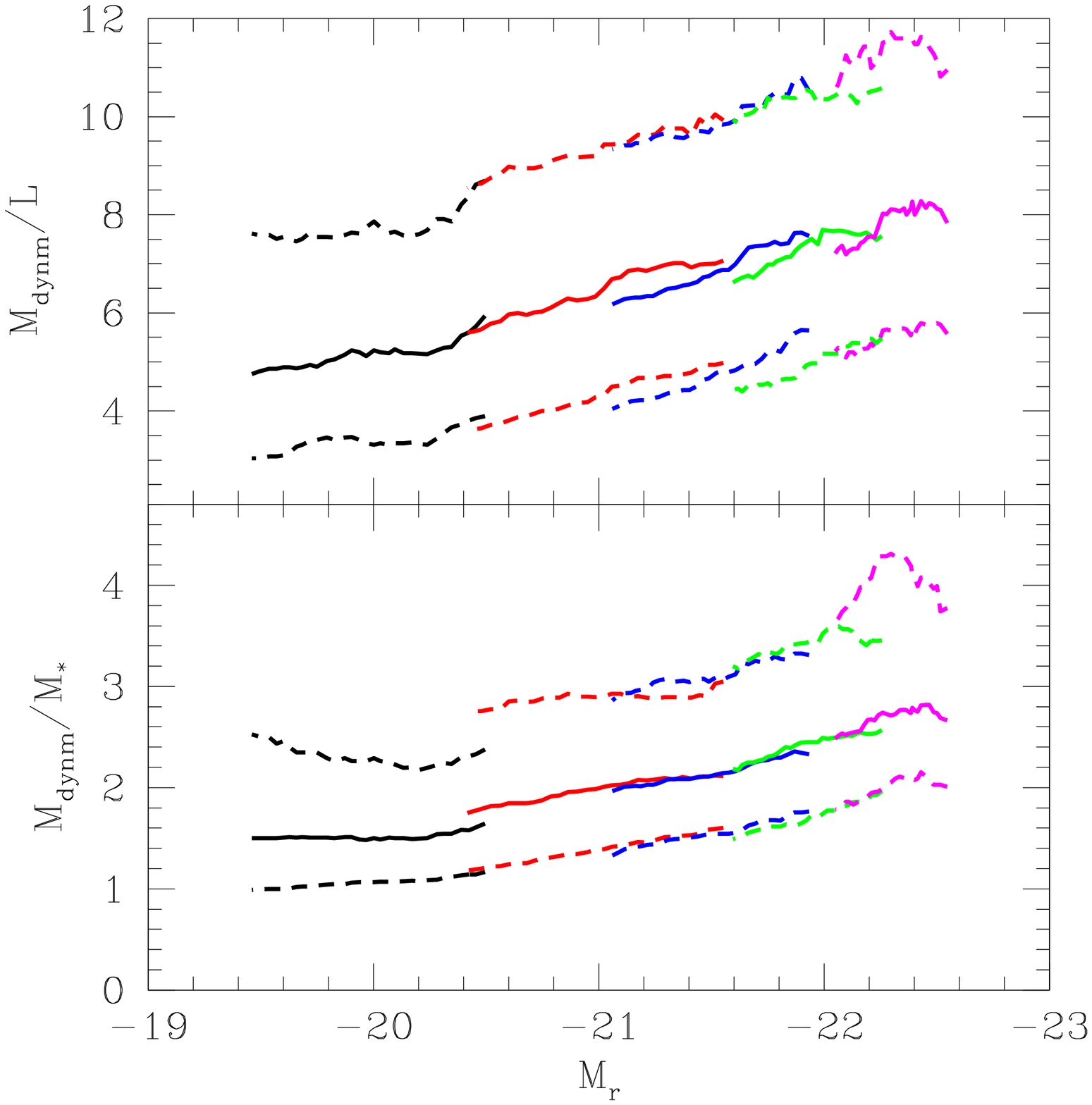}
\end{center}
\caption{ Same as Fig. \ref{fig:logml_l} except that the dynamical mass to light 
($M_{dynm}/L$) and the dynamical mass to stellar mass ratio ($M_{dynm}/M_{*}$) are 
plotted. Although we don't show it explicitly here, the scatter is consistent with 
log-normal.
}
\label{fig:logmdynl_l}
\end{figure}

Now that we have both stellar and dynamical masses, we can compare them.
Figs. \ref{fig:logml_l} and \ref{fig:logmdynl_l} show the stellar and dynamical mass
to light ratios respectively as a function of luminosity. 
A useful consistency test of both
the stellar and dynamical masses is that $M_{dynm}/L$ ($\sim 5-8$) is greater
than $M_{*}/L$ ($\sim 3$). Since the stellar masses are
sensitive to the choice of IMF, this constrains the possible
IMF's as we discuss in the next section. Also, $M_{dynm}/L$ increases roughly as $L^{0.17}$,
 while $M_{*}/L$ is constant, or slightly decreasing. 
The apparent decrease in $M_{*}/L$ 
with luminosity could be interpreted as luminosity evolution, although the 
statistical significance of the decrease is unclear (see below). 
Note that the scatter
in both quantities is consistent with being log-normal, with a width of $\sim 30$\%
for the stellar masses and slightly more for the dynamical masses.

A relevant question is whether this scatter is due to measurement error or 
an intrinsic scatter in galaxy properties. The errors in the stellar masses 
are obtained by integrating over a grid of population synthesis models to obtain 
the 16\% and 84\% confidence intervals; these errors are $\sim$ 25\%. Estimating 
an error for the dynamical mass is more involved; we assume a 10\% error 
on the conversion of $\sigma$ to the circular velocity (see Fig.\ref{fig:model} 
for a justification) and a further $\sim$ 10\% error on the measurement of 
$\sigma^{2} R_{50}$. This yields an error in the dynamical mass of $\sim 
30$ \%. This implies that the dominant source of scatter is measurement 
errors, so it is possible
that the intrinsic correlations are tighter than what is observed. 

The measurement errors can be dominated by systematic or statistical errors. 
Systematic errors can be caused by the velocity dispersion-mass conversion
we have assumed, errors
in extrapolation to $R_{50}$,
or by errors in the stellar mass determination. 
Pure statistical errors, such as those from measurement errors, will average out 
in median quantities. 
If statistical errors 
dominate, then the trends in the medians of these distributions
are highly statistically significant. 
If the errors are dominated by systematics, the trends may still be 
significant, as long as the systematic error does not couple to the parameter 
that is varied. In all our figures, we have presented the average $1\sigma$
contours for individual objects. This is the most conservative error estimate and 
it should be kept in mind that the errors on median quantities may be significantly 
smaller than this. 

\begin{figure}
\begin{center}
\leavevmode
\includegraphics[width=3.0in]{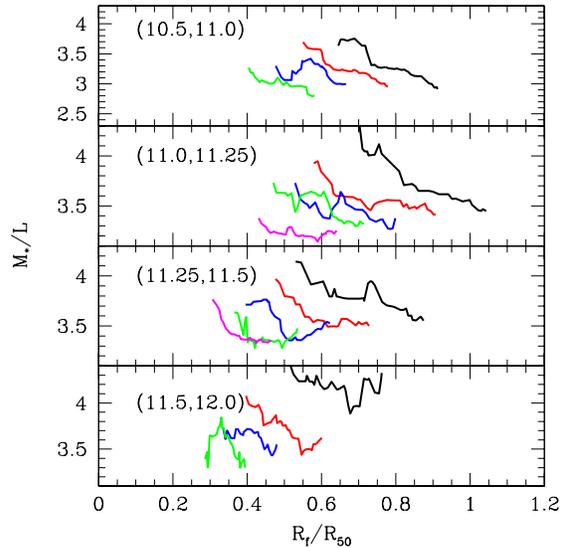}
\end{center}
\caption{  $M_{*}/L$ profiles for our sample, analogous to 
Fig.\ref{fig:velprof_plot}. The mass subsamples are labelled 
in each panel, while the individual lines are medians for
different physical sizes.
}
\label{fig:colourprof_plot}
\end{figure}

However, there are possible uncertainties in the stellar mass estimations 
that could lead to systematic errors. The stellar masses are estimated 
by comparing the measured $D_{4000}$ and $H\delta_{A}$ indices
with detailed stellar population synthesis models. In order to estimate the
robustness of the masses to these details, we compared the stellar masses used 
here with those computed by \cite{2002astro.ph.11546P} and found that, 
for the subsample of ellipticals found in both catalogs, the differences
were consistent with the errors. Specifically, all the trends
found in this paper are also present for the stellar masses estimated
by \cite{2002astro.ph.11546P}.

Another potential source of systematic error is colour gradients: $M_*/L$
is determined from the spectra within the fiber, typically at 0.4-0.8$R_{50}$,
and then assumed to hold within
$R_{50}$. We can try to measure an average $M_{*}/L$ profile using the same
method we used to estimate the velocity dispersion profile. This is shown 
in Fig.\ref{fig:colourprof_plot}; as before, the different panels are
different stellar mass ranges, while the different line segments are the median
$M_{*}/L$ for galaxies of similar physical sizes.
There is a  modest trend towards decreasing $M_{*}/L$
with radius, although the statistical significance is unclear.
If we attribute the observed trends to a colour gradient,
our stellar masses are only upper limits on the true stellar mass. 
The suggested trends would reduce the stellar mass by at most 20\%. 
While this would alter the precise form of the correlations that we have 
observed, it would increase the dark matter fraction observed, but would not
change any of the trends we find.
However, luminosity evolution would also cause a similar 
trend and is difficult to disentangle from $M_{*}/L$ gradients. 
Therefore, we choose to be conservative and assume that the estimated 
$M_{*}/L$ value is applicable at $R_{50}$. 
As a final check on systematics, we find that our trends remain unchanged 
even if we select only 
very spherical objects with $a/b>0.9$ that are unlikely to show any rotation.
This provides additional support to our 
assumption of ignoring rotation in the analysis. 

\begin{figure}
\begin{center}
\leavevmode
\includegraphics[width=3.0in]{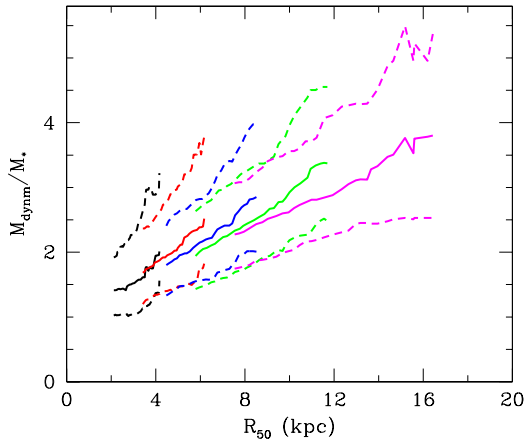}
\end{center}
\caption{ $M_{dynm}/M_{*}$ as a function of the physical size (half light radius) of the
galaxy. The different segments/colours correspond to cutting the sample in redshift as in 
Fig. \ref{fig:logml_l}, and then selecting a narrow luminosity bin around the median 
luminosity of the redshift subsample. The median luminosities for the subsamples are 
(from left to right) are -19.93, -20.78, -21.30, -21.78 and -22.20.
}
\label{fig:logml_LR}
\end{figure}

\begin{figure}
\begin{center}
\leavevmode
\includegraphics[width=3.0in]{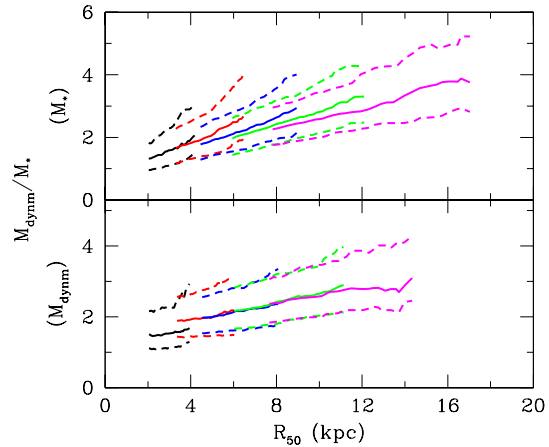}
\end{center}
\caption{ Same as Fig.\ref{fig:logml_LR} except that the subsample is cut in stellar
(upper panel) and dynamical (lower panel) mass. From left to right,
the median stellar masses are
4.36, 9.33, 15.3, 22.8 and 31.8 $\times 10^{10} M_{\odot}$, while the median
dynamical masses are 3.57, 9.51, 16.5, 27.1, 42.2 $\times 10^{10} M_{\odot}$.
}
\label{fig:logml_mR}
\end{figure}

We can now consider galaxies of a given luminosity and ask if there is any variation
in $M_{dynm}/M_{*}$ as a function of the size of the galaxy. 
In order to do this, we start with the volume limited samples discussed above and 
construct luminosity subsamples 0.5 magnitudes in width, centred about the median 
luminosity of each original sample.
Fig. \ref{fig:logml_LR}
shows $M_{dynm}/M_{*}$ as a function of $R_{50}$, for samples created in this manner. 
The ratio
of dynamical mass to stellar mass increases roughly linearly with radius, with a slope 
of $\sim 0.14$. Fig. \ref{fig:logml_mR} shows the same relation for stellar and 
dynamical mass subsamples; the relations for these subsamples are very similar to those
for luminosity.

It is interesting to ask whether these correlations can be explained with the 
model of Section \ref{sec:model}. The model has four parameters, the virial mass
and scale size of the halo, and the stellar mass and scale size of the galaxy. The 
latter two of these are directly constrained 
by observations. We parametrise the scale size
of the halo (Eq.\ref{eq:NFW}) by the concentration
parameter, $c \equiv r_{vir}/r_{s}$, where the virial radius, $r_{vir}$, is 
determined by the virial mass,
\be
M_{vir} = \frac{4\pi}{3}\Delta\rho_{crit}r_{vir}^{3} \,\,\,,
\ee
where $\Delta(=200)$ is the spherical overdensity and $\rho_{crit}$
is the critical density.  While numerical simulations suggest 
that the concentration decreases with virial mass, this variation is small over the 
range considered here. 
We thus make the assumption that the concentration is 
constant ($c=12.7$)
independent of the size of the halo.
This leaves us with only the virial 
mass of the halo to vary; Fig. \ref{fig:mass} compares the prediction of this 
model for the dynamical to stellar mass ratio as a function of $R_{50}$
with the observed result, for galaxies with stellar masses $\sim 1.5 \times
10^{11} M_{\odot}$. The model does well predicting the general trend for this
and other stellar masses, although the
slopes are shallower than 
what is observed. While the statistical significance of the 
disrepancy is small, it is possible that it is caused by the 
restrictive assumptions of the model.  
Allowing either the 
concentration or virial mass 
to vary with the size of the galaxy of a given stellar mass 
can reproduce the observed slope.

This simple one-parameter family of models provides a novel method
of estimating the virial mass of the halo in which a galaxy of a given stellar
mass resides. This method
is demonstrated in Fig.\ref{fig:mass}; one adjusts the virial mass to
best match the observed correlation of $M_{dynm}/M_{*}$ with $R_{50}$, 
while ``errors'' can be estimated by 
matching to the 16\% and 84\% contours. 
The results of this exercise are in 
Fig.\ref{fig:virialmass} as a function of stellar mass.
Here again we take the 
conservative approach of assuming the errors to be dominated by systematics. 
If the errors are statistical, then the usual $1/\sqrt{N}$ factor can be 
applied and the errors on the 
mean relation become considerably smaller. 
This figure 
argues that the virial masses of halos are
10 times the stellar mass
of ellipticals at the high mass end and somewhat less at the lower mass end. 

\section{Discussion}

\begin{figure}
\begin{center}
\leavevmode
\includegraphics[width=3.0in]{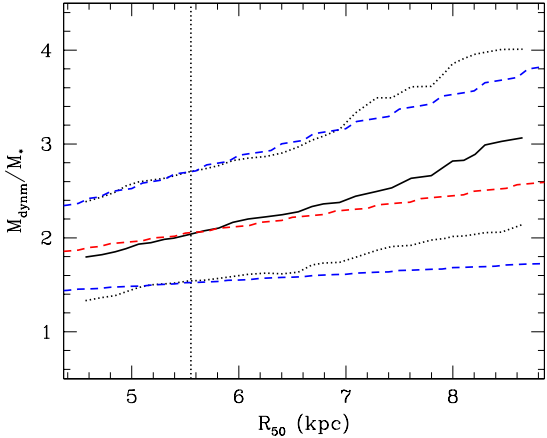}
\end{center}
\caption{The solid and dotted lines show the median and 16\% and 84\%
contours of the distribution of $M_{dynm}/M_{*}$ as a function of $R_{50}$
for a volume limited subsample with stellar mass between 1.17 and 1.86 
$\times 10^{11} M_{\odot}$ (median value of $1.5 \times 10^{11} M_{\odot}$).
The central dashed line shows the prediction of the model in Sec.\ref{sec:model} for the 
median stellar mass and a halo mass of $1.35 \times 10^{12} M_{\odot}$, while
the upper and lower lines have halo masses of $6.0 \times 10^{11} M_{\odot}$ (lower)
and $4.8 \times 10^{12} M_{\odot}$. All the initial halos had concentration parameters $c=12.7$.
The vertical dotted line is the median size of all the galaxies in the sample.
}
\label{fig:mass}
\end{figure}

\begin{figure}
\begin{center}
\leavevmode
\includegraphics[width=3.0in]{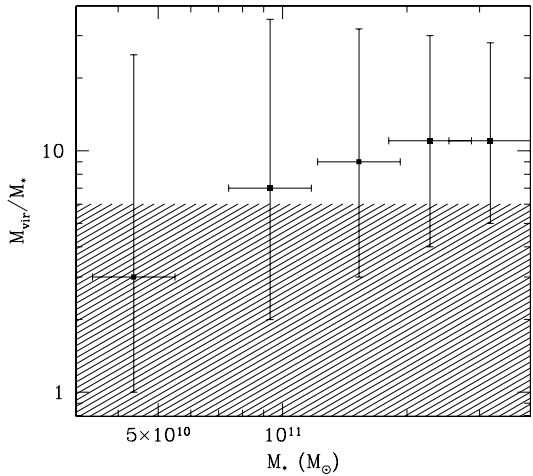}
\end{center}
\caption{ The ratio of the virial mass to stellar mass as a function of
stellar mass calculated using the methods described in the text and
Fig.\ref{fig:mass}. The shaded 
region is excluded by the cosmic baryonic fraction of $\sim 17$ \% 
measured by WMAP, assuming only baryons within the virial radius can 
cool and form stars.
}
\label{fig:virialmass}
\end{figure}

In this paper, we have compared stellar and dynamical mass estimates for 
a sample of elliptical galaxies in the SDSS, and studied the
dependence of that relation on the luminosity and size of the galaxy.

The first conclusion that we draw from this work is that stellar
mass estimates are consistent with the dynamical estimates, i.e. $M_{*} 
\leq M_{dynm}$ with $M_{*} \sim M_{dynm}$ for high surface brightness (low $R_{50}$) galaxies. 
This is far from a trivial conclusion, since the 
procedure to estimate stellar masses is a rather involved one. In particular,
the stellar masses are extremely sensitive to the stellar initial mass function.
Here we have followed \cite{2003MNRAS.341...33K} and adopt the 
\cite{2001MNRAS.322..231K} IMF; replacing
it with a Salpeter IMF increases the number of low-mass
stars, effectively increasing all the 
inferred stellar masses.
An examination of Fig. \ref{fig:logml_LR} shows that the stellar 
masses would no longer be consistent with the dynamical estimates in that case 
for high surface brightness (low $R_{50}$) galaxies.
This statement depends on how one regularises the divergence
of the Salpeter IMF at low masses, and therefore, must only be 
interpreted as a constraint on the minimum mass that the Salpeter slope 
can extend to. Our results suggest that a cutoff at $M>0.1M_{\sun}$ is required,
consistent with IMF measurements from young embedded clusters \citep{2002ApJ...573..366M}.

In what follows, we assume that the difference between
the dynamical and stellar masses is 
due to the presence of a dark matter component of the galaxies.
Ellipticals often have a hot gas component; 
however, it is a relatively minor contribution to the mass \citep{1997neg..conf..375S},
and the stellar mass is, to a good approximation, the baryonic mass of the 
galaxy.

Given that approximation, we can draw the following conclusions about
elliptical galaxies in SDSS :
\begin{itemize}
\item The fraction of dark matter increases with increasing luminosity: 
As is evident from Figs.\ref{fig:logml_l} and \ref{fig:logmdynl_l}, the 
stellar mass to light ratio is approximately independent of luminosity,
while $M_{dynm}/L \propto L^{0.17}$ , implying an increasing dark
matter fraction with luminosity. Note also that the dependence
of $M_{dynm}/L$ on $L$ is consistent with the results of \cite{2003AJ....125.1849B},
$M_{dynm}/L \propto L^{0.14 \pm 0.02}$. 

This result suggests that the tilt in the fundamental plane is caused by
an increasing dark matter fraction with luminosity. Recall that
the virial prediction for the fundamental plane,
\be
\sigma^{2} \propto \left( \frac{M}{L} \right) R \left(
\frac{L}{R^{2}} \right) \propto R I \,\, ,
\ee
is traditionally modified by assuming that the
mass to light ratio varies with luminosity, $(M/L) \propto L^{\alpha}$.
As discussed in the introduction, there are two possibilities
to explain this scaling, variations in metallicity or in the dark matter
fraction. Variations in metallicity would show up as variations in $M_{*}/L$,
while variations in the dark matter fraction would be evident in
$M_{dynm}/L$. We observe that $M_{*}/L$ is approximately independent of $L$,
while $M_{dynm}/L$ increases with $L$, favouring the conclusion that
the tilt is due to dark matter and not metallicity.  This is the
opposite conclusion from that reached by \cite{2001AJ....121.1936G}
who argue that a maximal stellar mass is supported by population
synthesis models, and therefore, one need not invoke dark matter.
Our approach is different; we use stellar masses estimated
independently of the dynamical estimates. Comparing these shows that
elliptical galaxies have a significant dark matter component within an 
effective radius, with as much as 3-4 times the stellar component
for the largest galaxies. 

We can also compare our results with those of \cite{2003MNRAS.341.1109B}, which attempted
to fit the fundamental plane with stellar and dark matter models similar
to those we use. However, \cite{2003MNRAS.341.1109B} leave the stellar mass to light ratio
a free parameter; this leads them to conclude that the data favours a 
model with no dark matter and $M_{*}/L \sim 5.3$. This is approximately 
a factor of 2 greater than our estimates of $M_{*}/L \sim 3$; moreover, we 
argue that changing our IMF to agree with the higher $M_{*}/L$ would 
make the stellar masses inconsistent with our dynamical estimates for
high surface brightness galaxies.

\item The dark matter fraction increases with increasing size at constant luminosity
(Fig. \ref{fig:logml_LR}). Note that this is a robust result, since similar trends
are also seen if the stellar mass or velocity dispersion are kept 
constant, instead of the luminosity (Fig.\ref{fig:logml_mR}). 
For the highest surface
brightness galaxies, the stellar mass approaches the dynamical mass. 
This suggests that the IMF based on local observations 
is also applicable to ellipticals. 
A simple model of a galaxy 
embedded in a dark matter halo with a constant star formation efficiency
\footnote{In other words, that the stellar mass fraction, $F$, is independent of 
the size of the galaxy.} 
captures the qualitative trends. In such a model a low surface 
brightness galaxy and a high surface
brightness galaxy of equal luminosity sit in equal dark matter halos, 
but the low surface brightness galaxy is more spread out, so the ratio of
stellar to dark matter mass is lower. 
These trends are thus expected in ab-initio models of galaxy formation. 
To explain them without the dark matter would require introducing an 
IMF that depends on the surface brightness. 

While this model can explain most, if not all, of the trends suggested 
by the observed relations, one question remains: what sets the scale 
size of the galaxy? As Fig. \ref{fig:logml_LR} shows, the half light radius varies 
by  a factor of two at any given luminosity. One simple possibility 
is that it is related to the scale radius of dark matter, 
which simulations suggest has a considerable scatter \citep{2001MNRAS.321..559B}. However, 
if a more compact galaxy is in a more concentrated dark matter halo 
then this would make the agreement in Fig. \ref{fig:mass} worse, since that 
figure suggests we need less dark matter in compact  
galaxies and more in low surface brightness galaxies.  In this case
the assumption of star formation efficiency being independent 
of galaxy scale size must also be violated. 
While we do not have definitive answers to these questions, we note that
ongoing galaxy-galaxy lensing studies on the same SDSS sample 
may be able to provide useful additional information,
since it can determine the virial 
mass as a function of luminosity, surface brightness etc. 
\end{itemize}

We also use this model to estimate the virial mass as a function of mass
and luminosity for our sample of galaxies. Although the errorbars are large,
we find that the virial mass of the halo is around 10 times the 
stellar mass of the galaxy (Fig. \ref{fig:virialmass}). We note that our 
modelling assumes that halos are undisturbed. Many of the ellipticals are in 
denser enviroments of groups and clusters. If the galaxy is at the center 
of the group then our modelling still applies. If it is a satellite then 
some of the dark matter attached to the galaxy before it merged into the 
group halo was likely stripped off due to the tidal effects. In this 
case our estimates apply to the virial mass of the galaxy prior to 
merging. This is the mass of interest if most of the stellar mass 
has been assembled prior to the galaxy merging into the group or cluster.

Comparing 10\% stellar mass fraction to the cosmic baryonic 
fraction of 17\% from WMAP \citep{2003astro.ph..2209S} implies 
that 60\% of the initial gas has been converted to stars; a $1\sigma$
uncertainty takes us between 25\% and 100\%.
This is not necessarily inconsistent 
with the global stellar mass fraction, which is only around 5-10\% of 
baryons \citep{2001MNRAS.326.1228B,2003ApJ...587...25D}: it suggests that 
the efficiency of star formation peaks 
at $M_{\rm vir}\sim 10^{12}M_{\sun}$
and is likely to be much smaller in either lower and higher mass halos.  
We observe this trend at the high mass end, where 
the dark matter fraction increases with increasing mass, although the
uncertainties in our measurement are large. 
The implication of an increasing dark matter fraction with
luminosity is that the star formation is suppressed in the most massive
halos, as expected for example by the longer cooling times in hotter halos. 
The sample of low mass ellipticals is too small to observe the reverse 
trend at the low mass end. 
Both the absolute value of the star formation efficiency and its trend 
with luminosity/stellar mass are in agreement with
galaxy-galaxy lensing results, which provide 
an independent estimate of the total mass of the halo 
\citep{2002MNRAS.335..311G}. 
These additional constraints on the mass also 
allow for a more detailed modelling 
of the structure of ellipticals. Ultimately, we hope that these 
results will both inform and constrain theories of galaxy formation.

We would like to thank Daniel Eisenstein, David Hogg, Raul Jimenez, 
Yeong-Shang Loh and Scott Tremaine for
useful discussions. U.S. acknowledges support from the Packard and Sloan 
foundations and NSF CAREER-0132953. M.A.S. acknowledges
support of NSF grant AST-0071091.

Funding for the creation and distribution of the SDSS Archive has been
 provided by the Alfred P. Sloan Foundation, the Participating Institutions,
 the National Aeronautics and Space Administration, the National Science Foundation,
 the U.S. Department of Energy, the Japanese Monbukagakusho, and the Max Planck
 Society. The SDSS Web site is http://www.sdss.org/. 

The SDSS is managed by the Astrophysical Research Consortium (ARC) for
 the Participating Institutions. The Participating Institutions are
 The University of Chicago, Fermilab, the Institute for Advanced
 Study, the Japan Participation Group, The Johns Hopkins University, Los Alamos National
 Laboratory, the Max-Planck-Institute for Astronomy (MPIA), the Max-Planck-Institute
 for Astrophysics (MPA), New Mexico State University, University of Pittsburgh, Princeton
 University, the United States Naval Observatory, and the University of Washington.

\appendix

\section{Properties of Sersic Profiles}
The Sersic profile is an obvious generalisation of the classification of galaxies
into deVaucouleurs and exponential intensity profiles,
\be
I(R) \propto \exp (-x^{1/n}) \,\, ,
\ee
where $x$ is a dimensionless radial variable, $R/R_{s}$. Exponential and deVaucouleurs 
profiles correspond to $n = 1$ and $n = 4$ respectively; in general, the Sersic
index $n$ can be any positive real number. Below, we summarise the relevant properties 
of Sersic profiles; a more detailed discussion is contained in \cite{1991A&A...249...99C}
and references therein.

The first quantity of interest is the half light radius
$R_{50}$.
Defining $y \equiv x^{1/n}$, we can solve for the integrated
intensity profile,
\bea
L(R) & \propto \int_{0}^{y} \, dy'\, {y'}^{2n - 1} \, \exp(-y') \,\, \nonumber \\
& \propto \gamma(2n,y) \,\,\, ,
\eea
where $\gamma(m,y)$ is the incomplete Gamma function. Using the above result,
we see that $R_{50}$ can be determined by solving,
\be
\frac{\gamma(2n,b(n))}{\Gamma(2n)} = \frac{1}{2} \,\, ,
\ee
where $\Gamma(2n) \equiv \gamma(2n,\infty)$ is the Gamma function and $b(n) = (R_{50}/R_{s})^{1/n}$. 
This equation 
can be solved numerically, but a useful approximation is $b(n) = 2n - 0.324$. 

\begin{figure}
\begin{center}
\leavevmode
\includegraphics[width=3.0in]{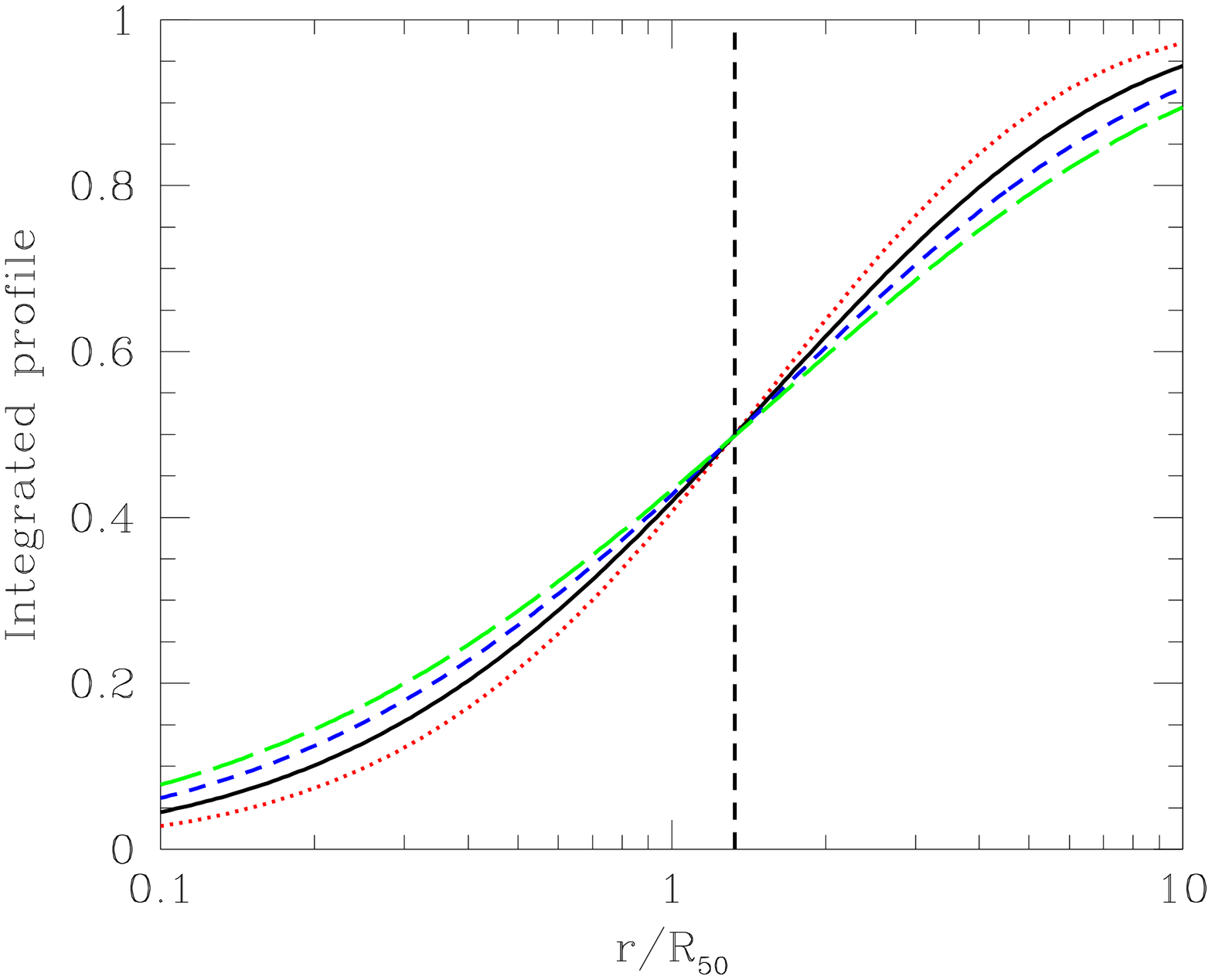}
\end{center}
\caption{Integrated light profiles for deprojected Sersic profiles for n=3 [dotted, red], 4 [solid, black],
5 [short-dashed, blue], 6 [long-dashed, green] profiles, as a function of $r/R_{50}$ where $R_{50}$ is the
projected half-light radius. The vertical dashed line marks the 3D half light radius for the n=4 profile.
}
\label{fig:sersic}
\end{figure}

We can also solve for the  deprojected Sersic profile $\nu(r)$; this is 
obtained via the Abel transform,
\be
\nu(r) = -\frac{1}{\pi} \int_{r}^{\infty} \, \frac{dI}{dR} \, \frac{dR}{\sqrt{R^{2} - r^{2}}} \,\,\, .
\ee
In order to quantify the differences between the different values of $n$, it is useful
to consider the total mass/light contained within a radius $r$. This is shown in 
Fig.\ref{fig:sersic} where we have scaled all the radii by $R_{50}$.
Note that
the discrepancies between different values of $n$ are marginal, especially at radii greater 
than $0.3 R_{50}$ (which contains less than 20\% of the total mass/light). This justifies 
our modelling the stellar light distribution of 
all galaxies with a Hernquist profile, even when $n \neq 4$. 

\bibliography{biblio,preprints}   
\bibliographystyle{mnras}

\end{document}